\documentclass[format=acmsmall, review=false]{acmart}

\usepackage{acm20} 
\usepackage{graphicx} 
\usepackage{algorithm}
\usepackage[noend]{algpseudocode}

\usepackage[T1]{fontenc}
\usepackage{graphicx} 
\usepackage{algorithm}
\usepackage[noend]{algpseudocode}
\usepackage{amsfonts}

\usepackage[utf8]{inputenc}
\usepackage{amsmath}
\usepackage{array}
\usepackage{graphicx}
\usepackage[english]{babel}
\usepackage[utf8]{inputenc}
\usepackage{algorithm}
\usepackage[noend]{algpseudocode}

\usepackage{enumitem}
\usepackage{wasysym}
\usepackage{framed}
\usepackage{pgfgantt,rotating}
\usepackage{subfloat}
\usepackage{multirow}
\usepackage{blindtext}

\begin{document}
\title{Frosty for partial synchrony}
%


\author{Stephen Buttolph}
\affiliation{%
\institution{ \institution{Ava Labs}  \city{NY} \country{USA}}
}

\author{Andrew Lewis-Pye}
\affiliation{%
\institution{ \institution{London School of Economics}  \city{London} \country{UK}}
}


\author{Kevin Sekniqi}
\affiliation{%
\institution{ \institution{Ava Labs}  \city{SF} \country{USA}}
}

\begin{abstract}
    Snowman is the consensus protocol used by blockchains on Avalanche.  Recent work has shown both how to augment Snowman with a `liveness' module called `Frosty' \cite{buchwald2024frosty} that protects against liveness attacks, and also how to modify Snowman so as to be consistent in partial synchrony \cite{buchwald2025snowman}. Since Frosty assumes (a strong form of) synchrony, the aim of this note is to show how to modify Frosty to deal with the partially synchronous version of Snowman described in \cite{buchwald2025snowman}. 
\end{abstract}

\maketitle              
\section{Introduction} \label{intro}

A version of Snowman for the synchronous setting  was first formally described and analysed in \cite{buchwald2024frosty}.  Since Snowman is subject to liveness attacks for adversaries larger than $O(\sqrt{n})$ (where $n$ is the number of validators/processes), that paper also described a `liveness module' that can be appended to the protocol to deal with attacks on liveness by  adversaries of size $O(n)$. The basic idea behind the liveness module is that one can use Snowman to reach fast consensus (with low communication cost) under normal operation, and can then trigger an `epoch change' that temporarily implements some standard quorum-based protocol to achieve liveness in the case that a substantial adversary attacks liveness. In the (presumably rare) event that a substantial adversary attacks liveness, liveness is thus ensured by \emph{temporarily} forgoing the communication complexity advantages of Snowman during normal operation. 

\vspace{0.2cm} 
When using the liveness module to defend against attacks by a large adversary, we therefore consider instructions that are divided into \emph{epochs}. In the first epoch (epoch 0), processes implement Snowman. In the event of a liveness attack, processes then enter epoch 1 and implement the quorum-based protocol to finalize a predetermined number of blocks. Once this is achieved, they enter epoch 2 and revert to Snowman, and so on. Processes only enter each odd epoch and start implementing the quorum-based protocol if a liveness attack during the previous epoch forces them to do so. The approach taken is reminiscent of protocols such as Jolteon and Ditto \cite{gelashvili2022jolteon}, in the sense that a view/epoch change mechanism is used to move between an optimistic and fallback path.

\vspace{0.2cm} 
More recently \cite{buchwald2025snowman}, it has been shown that a modification of Snowman called Snowman$^{\diamond}$ satisfies consistency in partial synchrony. However, the following issues remain: 
\begin{enumerate} 
\item Snowman$^{\diamond}$ is still subject to liveness attacks by  adversaries larger than $O(\sqrt{n})$, and;
\item  The analysis of \cite{buchwald2024frosty} only describes how to append a liveness module that integrates successfully with Snowman in the synchronous setting. 
\end{enumerate} 
The aim of this note is therefore to show how to append Snowman$^{\diamond}$  with a liveness model that functions correctly in partial synchrony. In what follows, we assume complete familiarity with \cite{buchwald2024frosty} and \cite{buchwald2025snowman}. In particular, we assume that the reader is familiar with Snowman$^{\diamond}$.

\section{The model} \label{model}

We consider a set $\Pi= \{ p_0,\dots, p_{n-1} \}$ of $n$ processes. Process $p_i$ is told $i$ as part of its input. For the sake of simplicity, we assume a static adversary that controls up to $f$ processes, where $f<n/5$. The bound $f<n/5$ is inherited from \cite{buchwald2025snowman}, where it was chosen only so as to make the consistency proofs for Snowman$^\diamond$ as simple as possible. 
 A process that is controlled by the adversary is referred to as \emph{Byzantine}, while processes that are not Byzantine are \emph{correct}. Byzantine processes may display arbitrary behaviour, modulo our cryptographic assumptions (described below).

\vspace{0.2cm} 
\noindent \textbf{Cryptographic assumptions}. Our cryptographic assumptions are standard for papers in distributed computing. Processes communicate by point-to-point authenticated channels. We use a cryptographic signature scheme, a public key infrastructure (PKI) to validate signatures, and a collision-resistant hash function $H$. 
 We assume a computationally bounded adversary. Following a common standard in distributed computing and for simplicity of presentation (to avoid the analysis of certain negligible error probabilities), we assume these cryptographic schemes are perfect, i.e.\ we restrict attention to executions in which the adversary is unable to break these cryptographic schemes. In a given execution of the protocol, hash values are therefore assumed to be unique.

 \vspace{0.2cm} 
\noindent \textbf{Communication and clocks}. As noted above, processes communicate using point-to-point authenticated channels. We consider the standard  partially synchronous setting:  For some known $\Delta$ and some unknown \emph{Global-Stabilization-Time} (GST), a message sent at time $t$ must arrive by time $\text{max} \{ t, \text{GST} \} +\Delta$. This setting formalizes the requirement that consistency must be maintained in asynchrony (when there is no bound on message delays), but liveness need only be ensured given sufficiently long intervals of synchrony.

\vspace{0.1cm}
We do not suppose that the clocks of correct processes are synchronized. For the sake of simplicity, we do suppose that the clocks of correct processes all proceed in real time, i.e.\ if $t'>t$ then the local clock of correct $p$ at time $t'$ is $t'-t$ in advance of its value at time $t$. This assumption is made only for the sake of simplicity, and our arguments are easily adapted to deal with a setting in which there is a given upper bound on the difference between the clock speeds of correct processes. 


 \vspace{0.2cm} 
\noindent \textbf{The binomial distribution}. Consider $k$ independent and identically distributed random variables, each of which has probability $x$ of taking the value `red'. We let $\text{Bin}(k,x,m)$ denote the probability that $m$ of the $k$ values are red, and we write $\text{Bin}(k,x,\geq m)$ to denote the probability that \emph{at least} $m$ values are red (and similarly for $\text{Bin}(k,x,\leq m)$).

 \vspace{0.2cm} 
\noindent \textbf{Dealing with small probabilities}. In analysing the security of a cryptographic protocol, one standardly regards a function $f:\mathbb{N} \rightarrow \mathbb{N}$ as \emph{negligible} if, for every $c\in \mathbb{N}$, there exists $N_c\in \mathbb{N}$ such that, for all $x\geq N_c$, $|f(x)|<1/x^c$. Our concerns here, however, are somewhat different. As noted above, we assume the cryptographic schemes utilized by our protocols are perfect. For certain \emph{fixed} parameter values, we want to be able to argue that error probabilities are sufficiently small that they can reasonably be dismissed. 
In our analysis, we will therefore identify certain events as occurring with \emph{small} probability (e.g. with probability $<10^{-20}$), and may then condition on those events not occurring. 
Often, we will consider specific events, such as the probability in a round-based protocol that a given process performs a certain action $x$ in a given round. In dismissing small error probabilities, one then has to take account of the fact that there may be many opportunities for an event of a given type to occur, e.g. any given process may perform action $x$ in any given round. How reasonable it is to condition on no correct process performing action $x$ may therefore depend on the number of processes and the number of rounds, and we assume `reasonable' bounds on these values. We will address such accountancy issues as they arise.

\section{Informal overview} 

In what follows, we assume familiarity with \cite{buchwald2024frosty} and \cite{buchwald2025snowman}.  Recall that the local variable $\mathtt{pref}$ is a process's presently preferred chain and that $\mathtt{final}$ is its presently finalized chain.  The instructions for Snoman$^{\diamond}$ are specified by a number of parameters: $k$ determines sample sizes, $\alpha_1$ and $\alpha_2$ are used to help determine a processes current $\mathtt{pref}$ values, while $\beta$ (as well as $\alpha_2$) are used to determine when a process should finalize new values. In what follows, we will focus on providing a formal analysis for the case that $k=80$, $\alpha_1=41$, $\alpha_2=72$ and $\beta=12$, but the analysis is easily modified to deal with alternative parameter values. 

\vspace{0.2cm} We first recall some aspects of the liveness module which remain unchanged from \cite{buchwald2024frosty}. 

\vspace{0.2cm} 
\noindent \textbf{The use of epochs}. As in \cite{buchwald2024frosty}, and as outlined in Section \ref{intro}, the basic idea is to run a version of the Snowman protocol called Snowman$^{\diamond}$ during standard operation, and to temporarily fall back to a standard `quorum-based' protocol in the event that a substantial adversary attacks liveness for Snowman$^{\diamond}$. We therefore consider instructions that are divided into \emph{epochs}. In the first epoch (epoch 0), processes implement Snowman$^{\diamond}$. In the event of a liveness attack, processes then enter epoch 1 and implement the quorum-based protocol to finalize a set number of new blocks. Once this is achieved, they enter epoch 2 and revert to Snowman$^{\diamond}$, and so on. Processes only enter each odd epoch and start implementing the quorum-based protocol if a liveness attack during the previous epoch forces them to do so. 

\vspace{0.2cm} 
\noindent \textbf{Adding a decision condition}. In even epochs, and when a process sees sufficiently many consecutive rounds during which its local value $\mathtt{final}$ remains unchanged, it will send a message to others indicating that it wishes to proceed to the next epoch. Before any correct process $p_i$ enters the next epoch, however, it requires messages from at least 1/5  of all processes indicating that they wish to do the same. This is necessary to avoid the adversary being able to trigger a change of epoch at will, but produces a difficulty:  some correct processes may wish to enter the next epoch, but the  number who wish to do so may not be enough to trigger the epoch change. To avoid such a situation persisting for an extended duration, we introduce (as in \cite{buchwald2024frosty}) an extra decision condition. Processes now report their value $\mathtt{final}$ as well as their value $\mathtt{pref}$ when sampled. We consider an extra parameter $\alpha_3$: for our analysis here,\footnote{Recall that we focus on the case $k=80$, $\alpha_1=41$, $\alpha_2=72$, $\beta=12$.} we suppose $\alpha_3=48$ (since $48=\frac{3}{5}\cdot 80$). If $p_i$ sees two consecutive samples in which at least $\alpha_3$ processes report $\mathtt{final}$ values that all extend $\sigma$, then $p_i$ will regard $\sigma$ as final. For $k=80$, $\alpha_3=48$ and if $f<n/5$, the probability that at least 3/5 of $p_i$'s sample sequence in a given round are Byzantine is less than $10^{-14}$, so the probability that this happens in two consecutive rounds is small. Except with small  probability, the new decision rule therefore only causes $p_i$ to finalize $\sigma$ in the case that a correct process has already finalized this value, meaning that it is safe for $p_i$ to do the same. Using this decision condition, we will be able to argue (exactly as in \cite{buchwald2024frosty}) that epoch changes are triggered in a timely fashion: either the epoch change is triggered soon after any correct $p_i$ wishes to change epoch, or else sufficiently many correct processes do not wish to trigger the change that $p_i$ is quickly able to finalize new values. 

\vspace{0.2cm} 
\noindent \textbf{Epoch certificates}. While in even epoch $e$, and for a parameter $\gamma$ (chosen to taste),\footnote{In Section \ref{clanalysis}, we suppose $\gamma\geq 300$. While we leave the technical conditions for entering an odd epoch the same as in \cite{buchwald2024frosty}, we note that the assumption of partial synchrony (rather than synchrony, as assumed in \cite{buchwald2024frosty}) now means that long periods of asynchrony (rather than an attack on liveness during synchrony) may now suffice to trigger the move to an odd epoch. Practically speaking, one may choose to ameliorate this by choosing $\gamma$ large enough that the move to an odd epoch could only be triggered by a period of asynchrony longer than is likely to occur in the absence of any deliberate attack on liveness. } $p_i$ will send the (signed) message $(\text{stuck},e,\mathtt{final})$ to all others when it sees $\gamma$ consecutive rounds during which its local value $\mathtt{final}$ remains unchanged. This message indicates that $p_i$ wishes to enter epoch $e+1$ and is referred to as an `epoch $e+1$ message'. For any fixed $\sigma$, a set of messages of size at least $n/5$, each signed by a different process and of the form $(\text{stuck},e,\sigma)$, is called an \emph{epoch certificate} (EC) for epoch $e+1$.\footnote{To ensure ECs are strings of constant bounded length (independent of $n$), one could use standard `threshold' cryptography techniques \cite{boneh2001short,shoup2000practical}, but we will not concern ourselves with such issues here.} When $p_i$ sees an EC for epoch $e+1$, it will send the EC to all others and enter epoch $e+1$. 

\vspace{0.2cm} Next, we describe some aspects of the liveness module which differ from \cite{buchwald2024frosty}. 

\vspace{0.2cm} 
\noindent \textbf{The use of Simplex}. In \cite{buchwald2024frosty}, a form of Tendermint was used during odd epochs. Here, we will use a form of Simplex \cite{chan2023simplex}, and we assume familiarity with that protocol. 

\vspace{0.2cm} 
\noindent \textbf{Ensuring consistency between epochs}. We must ensure that the values finalized by Simplex during an odd epoch $e+1$ extend all $\mathtt{final}$ values for correct processes. The basic intuition here is very similar to that in \cite{buchwald2024frosty}, but the use of Snowman$^\diamond$ rather than Snowman$^+$ impacts some details of the consistency analysis. As in  \cite{buchwald2024frosty}, the rough idea is that we have processes send out their local $\mathtt{pref}$ values upon entering epoch $e+1$, and then use these values to extract a chain that it is safe for the quorum based protocol (Simplex, in this case)  to build on. 
Upon entering epoch $e+1$, we therefore have $p_i$ send out the message $(\text{start},e+1,\mathtt{pref})$. 
 This message is referred to as a \emph{starting vote} for epoch $e+1$ and, for any string $\sigma$, we say that the starting vote $(\text{start},e+1,\mathtt{pref})$ extends $\sigma$ if $\sigma \subseteq \mathtt{pref}$.  By a \emph{starting certificate} (SC) for epoch $e+1$ we now mean a set of at least $4n/5$ starting votes for epoch $e+1$, each signed by a different process. If $C$ is an SC for epoch $e+1$, we set $\mathtt{Pref}(C)$ to be the longest $\sigma$ extended by more than half of the messages in $C$. The basic idea is that $\mathtt{Pref}(C)$ must extend all $\mathtt{final}$ values for correct processes, and that consistency will therefore be maintained if we have Simplex finalize a value extending this string. 

To argue that this is indeed the case, recall the proof described in Section 8 of  \cite{buchwald2025snowman} (and recall that $f<n/5$). The proof given there suffices to establish that, if any correct process $p_i$ finalizes $\sigma$ at some time $t$, then (except with small  error probability),  more than 3/4 of the correct processes have local $\mathtt{pref}$ values that extend $\sigma$ at all timeslots starting from some point prior to $t$ and prior to any point at which they enter the next odd epoch, and continuing until either the end of the protocol execution or any point at which they enter the next odd epoch. This  suffices to ensure that $\mathtt{Pref}(C)$ will extend $\sigma$: since $\frac{4}{5}\cdot \frac{3}{4}=\frac{3}{5}$, and since $S$ contains at least $4n/5$ starting votes,  more than half the votes in $C$ must extend $\sigma$.

%
%
%
%
%
%
%
%
%
%
%
%
%
%
%
%
%
%
%
%

\section{Frosty the Liveness Module: formal specification} 

During odd epochs, we run a protocol which is essentially just Simplex, modified slightly so that an SC is used to ensure consistency with the previous epoch. 


\vspace{0.1cm} 
When a correct process $p_i$ sends a message $m$ `to all', it is convenient to suppose that  $p_i$ regards $m$ as being sent to itself and `received' at the next timeslot. All sent messages are signed by the process sending the message. We consider the following local variables, procedures, and functions for $p_i$.

\vspace{0.1cm} 
\noindent $\mathtt{e}$: The epoch in which $p_i$ is presently participating (initially 0). 

\vspace{0.1cm} 
\noindent $\mathtt{ready}(e)$: Indicates whether $p_i$ has already initialized values for epoch $e$. Initially, $\mathtt{ready}(e)=0$.

\vspace{0.1cm} 
\noindent \textbf{Starting votes}: A starting vote for epoch $e$ is a message of the form 
$(\text{start},e,\sigma)$ for some $\sigma$. 
For any string $\sigma'$, we say that the starting vote $(\text{start},e,\sigma)$ extends $\sigma'$ if $\sigma' \subseteq \sigma$. 

\vspace{0.1cm} 
\noindent \textbf{Starting certificates}: A starting certificate for epoch $e$ is  set of at least $4n/5$ starting votes for epoch $e$, each signed by a different process.

\vspace{0.1cm} 
\noindent $\mathtt{Pref}(C)$: If $C$ is a starting certificate (SC) for epoch $e$, we set $\mathtt{Pref}(C)$ to be the longest $\sigma$ extended by more than half of the messages in $C$.

\vspace{0.1cm} 
\noindent $\mathtt{h}$:  the `view' in which $p_i$ is presently participating (upon entering an odd epoch, this is initially set to 1).

\vspace{0.1cm} 
\noindent \textbf{Blocks}. A Simplex-block $b$ for epoch $e$ is a tuple $(h, \mathtt{parent},\mathtt{txs},e,C)$, where:
\begin{itemize} 
\item $h\in \mathbb{N}_{\geq 1}$ is referred to as the \emph{height} of the block;
\item  $\mathtt{parent}$ is a string that (typically) is meant to be
the hash of a “parent” blockchain;
\item $\mathtt{txs}$ is a sequence of transactions;
\item $e$ indicates the epoch corresponding to the block, and; 
\item $C$ is a starting certificate for epoch $e$ (unless the block is a ``dummy block'', see below).
\end{itemize} 
The instructions build blocks, beginning with $h=1$ for each odd epoch. If $b=(h, \mathtt{parent},\mathtt{txs},e,C)$, then we set $b.\text{h}:=h$, $b.\text{parent}:=\mathtt{parent}$, $b.\text{txs}:=\mathtt{txs}$, $b.\text{e}:=e$, and $b.\text{C}:=C$. 

\vspace{0.1cm} 
\noindent \textbf{Dummy blocks}.  The special Simplex-dummy-block of height $h$ for epoch $e$ is the tuple $\bot_{e,h} := (h, \bot, \bot,e,\bot)$. As in the standard version of Simplex, this
is an empty block that will be inserted into the blockchain at heights where no agreement is
reached. A dummy block does not point to a specific parent.

\vspace{0.1cm} 
\noindent \textbf{Simplex Blockchains}. A Simplex-blockchain of height $h$ for epoch $e$ is a sequence of Simplex-blocks $(b_1,\dots, b_h)$ such that, for each $i \in [h]$:
\begin{itemize} 
\item Either $b_i=\bot_{e,i}$, or else $b_i=(i, H(b_1,\dots, b_{i-1}),\mathtt{txs},e,C)$ for some $\mathtt{txs}$ and some $C$ which is an SC for epoch $e$; 
\item If $b_i\neq \bot_{e,i}$ and there exists a greatest $j<i$ such that $b_j\neq \bot_{e,j}$ then $b_i.\text{C}=b_j.\text{C}$. 
\end{itemize}   

\vspace{0.1cm} 
\noindent \textbf{Notarized blocks and blockchains}.  These are defined essentially as in standard Simplex. A \emph{notarization} for a Simplex-block $b$ ($b$ may be the
dummy block) is a set of signed messages of the form $(\text{vote},e, h, b)$ from $>4n/5$ different
processes, where $h=b.\text{h}$ and $e=b.\text{e}$. A \emph{notarized Simplex-block} is a Simplex-block augmented
with a notarization for that block.

\vspace{0.1cm} 
\noindent A \emph{notarized Simplex-blockchain} for epoch $e$ is a tuple $(b_1,\dots, b_h, S)$, where $b_1,\dots , b_h$ is a Simplex-blockchain for epoch $e$, and $S$ is a
set of notarizations, one for each of $b_1,\dots, b_h$.  A notarized Simplex-blockchain may
contain notarized dummy blocks.

\vspace{0.1cm} 
\noindent \textbf{Finalized blocks and blockchains}.  A \emph{finalization} for  height $h$ in epoch $e$ is a set of signed messages
of the form $(\text{finalize}, e,h)$ from $>4n/5$ different processes. We say that a block of height
$h$ for epoch $e$ is finalized if it is notarized and accompanied by a finalization for $h$ in epoch $e$.
A \emph{finalized Simplex-blockchain for epoch $e$} is a notarized Simplex-blockchain for epoch $e$, accompanied
by a finalization for the last block.

\vspace{0.1cm} 
\noindent \textbf{The reduced height of a blockchain}. The \emph{reduced height}  of  a Simplex-blockchain $b_1,\dots,b_h$  for epoch $e$ is the number of $j\in [h]$ such that $b_j\neq \bot_{e,j}$. Let $j_1<\dots<j_{h'}$ be those  $j\in [h]$ such that $b_j\neq \bot_{e,j}$ (so that $h'$ is the reduced height of $b_1,\dots,b_h$). We set $\text{reduce}(b_1,\dots,b_h):=b_{j_1},\dots,b_{j_{h'}}$. 

\vspace{0.1cm} 
\noindent \textbf{The number of blocks finalized in odd epochs}. We consider a parameter $\mu\in \mathbb{N}_{\geq 1}$, chosen to taste, which plays a role in determining when we switch from an odd epoch to the next even epoch. If $b_1,\dots,b_h$ is a finalized Simplex-blockchain for epoch $e$ of reduced height $\geq \mu$, then let $b_{j_1},\dots,b_{j_{h'}}:=\text{reduce}(b_1,\dots,b_h)$ and let $C$ be such that $b_{j_i}.\text{C}=C $ for all $i\in [h']$. We set $\text{fin}(b_1,\dots,b_h):= \mathtt{Pref}(C) \ast H(b_{j_1}) \ast \dots \ast H(b_{j_{\mu}})$, where $\ast$ denotes concatenation. We will run odd epochs until producing a finalized Simplex-blockchain for the epoch of reduced height $\geq \mu$, and will then use the function $\text{fin}$ to determine the finalized sequence that correct processes carry through to the next even epoch.\footnote{A subtle point is that, while $\mathtt{Pref}(C)$ will be a concatenated sequence of block hashes, the final block in the corresponding sequence may have its hash only partially specified by $\mathtt{Pref}(C)$. This is not problematic, but requires consideration when interpreting which \emph{transactions} are specified as finalized by the sequence $\text{fin}(b_1,\dots,b_h)$. }


\vspace{0.1cm} 
\noindent \textbf{Leaders}. We let $\mathtt{lead}(h)$ denote the \emph{leader} for height $h$ while in an odd epoch. We set $\mathtt{lead}(h)=p_j$, where $j= h \text{ mod }n$.

\vspace{0.1cm} 
\noindent \textbf{The procedure $\mathtt{MakeProposal}$}. : If $p_i = \mathtt{lead}(h)$, then upon entering view $h$ while in epoch $e$,  $p$ sends to all  processes a single signed proposal of the form
\[ (\text{propose}, e,h, b_1,\dots , b_{h-1}, b_h, S).\] 
Here, $b_0,\dots, b_h$  is $p_i$'s choice of a Simplex-blockchain of height $h$ for epoch $e$, where $(b_1,\dots,b_{h-1}, S)$ is a notarized
Simplex-blockchain for epoch $e$,  and $b_h\neq \bot_{e,h}$. The block $b_h$ should contain every pending transaction
seen by $p_i$ but not already in the parent chain.

\vspace{0.2cm} 
\noindent \textbf{Conventions regarding the gossiping of blocks and other messages while in an odd epoch}. While in odd epochs, we suppose that correct processes automatically gossip received blocks, notarizations and finalizations. For example, this means that if $p_i$ is correct and sees a notarized blockchain of height $h$ for epoch $e$ at time $t$ after GST, then all correct processes will have seen those messages by  $t+\Delta$. 

\vspace{0.2cm} 
\noindent   The pseudocode appears in Algorithm 1.

\begin{algorithm} \label{pc:Ava1.5odde}
\caption{Frosty: The instructions for process $p_i$ $\textbf{while}\  \mathtt{e}$ is odd}
\begin{algorithmic}[1]

\State At every $t$ \textbf{if} $\mathtt{ready}(\mathtt{e})==0$:


  \State \hspace{0.3cm} Send $(\text{start},\mathtt{e},\mathtt{pref})$ to all if not already sent; \Comment{Send starting vote}

  \State \hspace{0.3cm} \textbf{If} $p_i$ has received an SC for epoch $\mathtt{e}$, set $\mathtt{ready}(\mathtt{e}):=1$, $\mathtt{h}:=1$;

    \State 
    
    \State At every $t$ \textbf{if} $\mathtt{ready}(\mathtt{e})==1$:

%
%
%
%

\State \hspace{0.3cm} Set timer $T_{\mathtt{e},\mathtt{h}}$ to fire in $3\Delta$ time, if not already done; \Comment Set timer 

\State 

  \State \hspace{0.3cm} \textbf{If} $p_i==\mathtt{lead}(\mathtt{h})$:     

    \State \hspace{0.6cm} $\mathtt{MakeProposal}$ if not already done;     \Comment Send proposal if leader

      \State 
      
      \State \hspace{0.3cm}  If $T_{\mathtt{e},\mathtt{h}}$ fires, send the signed message $(\text{vote},\mathtt{e}, \mathtt{h}, \bot_{\mathtt{e},\mathtt{h}})$ to all; \Comment Vote for dummy block

\State

\State \hspace{0.3cm} \textbf{Upon} receiving a \emph{first} proposal from $\mathtt{lead}(\mathtt{h})$ of the form $(\text{propose}, \mathtt{e},\mathtt{h}, b_1,\dots , b_{\mathtt{h}-1}, b_\mathtt{h}, S)$, 

\State \hspace{0.3cm} check to see that: 
\State \hspace{0.3cm} (i) $b_\mathtt{h}\neq \bot_{\mathtt{e},\mathtt{h}}$,
\State \hspace{0.3cm}  (ii) $b_1,\dots, b_\mathtt{h}$ is a valid Simplex-blockchain for epoch $\mathtt{e}$, and 
\State \hspace{0.3cm} (iii) $(b_1,\dots b_{\mathtt{h}-1},S)$ is a notarized  blockchain.
\State \hspace{0.3cm}  If so, send the signed message $(\text{vote},\mathtt{e}, \mathtt{h}, b_\mathtt{h})$ to all; \Comment Vote for block

\State 

\State \hspace{0.3cm} \textbf{Upon} seeing a notarized blockchain of height $\mathtt{h}$:

\State \hspace{0.6cm}  \textbf{If} $T_{\mathtt{e},\mathtt{h}}$ did not fire yet: 
\State \hspace{0.9cm} Cancel $T_{\mathtt{e},\mathtt{h}}$ (so that it never fires);
\State \hspace{0.9cm}  Send the signed message $(\text{finalize},\mathtt{e},\mathtt{h})$ to all;   \Comment Finalize a new block

\State \hspace{0.6cm} Set $\mathtt{h}:=\mathtt{h}+1$; 

\State

\State \hspace{0.3cm} \textbf{Upon} receiving a finalized Simplex-blockchain $b_1,\dots,b_{h'}$ for epoch $\mathtt{e}$ of reduced height 
\State \hspace{0.3cm} $\geq \mu$:         \Comment $\mu$ is parameter, set to taste
\State \hspace{0.6cm} Set $\mathtt{final}:=\text{fin}(b_1,\dots,b_{h'})$, $\mathtt{e}:=\mathtt{e}+1$;    \Comment Move to next epoch

\end{algorithmic}
\end{algorithm}

\section{Modifying Snowman$^\diamond$ to integrate with Frosty}

Once again, we assume familiarity with \cite{buchwald2024frosty} and \cite{buchwald2025snowman} in what follows. During even epochs, processes essentially just execute Snowman$^\diamond$ (as specified in  \cite{buchwald2025snowman}), with modifications (which are similar to those in \cite{buchwald2024frosty}) that allow for transition to the next  epoch. 

\vspace{0.2cm} 
\noindent \textbf{The variables, functions and procedures used by $p_i$}. The protocol instructions make use of the following variables and functions (as well as others whose use should be clear from the pseudocode):

\begin{itemize}
\item $b_0$: The genesis block.
\item $\mathtt{blocks}$: Stores blocks received by $p_i$ (and verified as valid). Initially it contains only $b_0$, and it is automatically updated over time to include any block included in any message received or sent by $p_i$.
\item $\mathtt{val}(\sigma)$, initially undefined: For each finite binary string $\sigma$, $\mathtt{val}(\sigma)$ records $p_i$'s presently preferred value for the next bit of the chain of hash values $H(b_0) \ast H(b_1) \ast \dots$, should the latter extend $\sigma$. 

\item $\mathtt{pref}$, initially set to $H(b_0)$: The initial segment of the chain of hash values that $p_i$ presently prefers. We write $|\mathtt{pref}|$ to denote the length of this binary string.
\item $\mathtt{pref}(s,e)$, initially undefined: $p_i$'s $\mathtt{pref}$ value at the end of round $s$ of (even) epoch $e$. 
\item $\mathtt{final}$, initially set to $H(b_0)$: The initial segment of the chain of hash values that $p_i$ presently regards as final. 

\item $\mathtt{chain}(\sigma)$: If there exists a greatest $h\in \mathbb{N}$ such that\footnote{The move from even to odd epochs may cause some block hash values to be partial: we subdue such issues here for the sake of simplicity.}  $\sigma=H(b_0) \ast \dots \ast H(b_h) \ast \tau$ for a chain of blocks $b_0 \ast \dots \ast b_h$ all seen by $p_i$, and for some finite string $\tau$, then $\mathtt{chain}(\sigma):= b_0 \ast \dots \ast b_h$. Otherwise, $\mathtt{chain}(\sigma):=b_0$. 

\item $\mathtt{reduct}(\sigma)$: If there exists a greatest $h\in \mathbb{N}$ such that $\sigma=H(b_0) \ast \dots \ast H(b_h) \ast \tau$ for a chain of blocks $b_0 \ast \dots \ast b_h$ all seen by $p_i$, and for some finite string $\tau$, then $\mathtt{reduct}(\sigma):=H(b_0) \ast \dots \ast H(b_h)$. Otherwise, $\mathtt{reduct}(\sigma):=H(b_0)$. 

\item $\mathtt{last}(\sigma)$: If there exists a greatest $h\in \mathbb{N}$ such that $\sigma=H(b_0) \ast \dots \ast H(b_h) \ast \tau$ for a chain  $b_0 \ast \dots \ast b_h$ all seen by $p_i$, and for some finite string $\tau$, then $\mathtt{last}(\sigma):= b_h$. Otherwise, $\mathtt{last}(\sigma):=b_0$.
\item $H_B$: If $B=b_0\ast b_1 \ast \dots \ast b_h$ is a chain, then $H_B:=H(b_0) \ast H(b_1) \ast \dots H(b_h)$, and if not then $H_B$ is the empty string $\emptyset$.
\item $\mathtt{start}(s,e)$, initially undefined: The time at which $p_i$ starts round $s$ of epoch $e$ (according to local clock).

\item  $\mathtt{lock}(\sigma)$, initially 0: Indicates whether $p_i$ is locked on $\sigma$.

\item $\mathtt{locktime}(\sigma)$, initially undefined: The time at which $p_i$ became locked on $\sigma$.

\item $\mathtt{lockbound}(\sigma)$, initially 0: Rounds after this may cause $p_i$ to become freshly locked.

\item $\mathtt{newround}$, initially 1: Indicates whether $p_i$ should start a new round.

\item  $\mathtt{s}$, initially 0: Present round. 

\item $\mathtt{Init}(e)$: This process is run at the beginning of even epoch $e$, and performs the following: Set $\mathtt{pref}:=\mathtt{final}$, $\mathtt{s}:=0$, $\mathtt{newround}:=1$,  and for all $\sigma$ set $\mathtt{lock}(\sigma):=0$, $\mathtt{lockbound}(\sigma):=0$,  and make $\mathtt{val}(\sigma)$ and $\mathtt{locktime}(\sigma)$ undefined. 

\item $\mathtt{suppfin}(\sigma,s,e)$, initially 0: Records whether responses from round $s$  of epoch $e$ support finalizing $\sigma$.

\item $\mathtt{dec}(s,\sigma,e)$, initially 0: Records whether values for round $s$ of epoch $e$ already suffice to determine that $\sigma$ is an initial segment of $\mathtt{pref}(s)$.

\item $\mathtt{rpref}(j,s,e)$, initially undefined: The preferred chain that $p_{j,s}$ reports to $p_i$ in round $s$ of epoch $e$. 

\item $\mathtt{rlock}(j,s,e)$, initially undefined: The initial segment of $\mathtt{rpref}(j,s,e)$ that $p_{j,s}$ reports to $p_i$ as having been locked for time $4\Delta$. 

\item $\mathtt{rfin}(j,s,e)$, initially undefined: The initial segment of $\mathtt{rpref}(j,s,e)$ that $p_{j,s}$ reports to $p_i$ as being finalized. 

\item For any fixed $\sigma$, a set of messages of size at least $n/5$, each signed by a different process and of the form $(\text{stuck},e,\sigma)$, is called an \emph{epoch certificate} (EC) for epoch $e+1$.
   
\end{itemize}

The pseudocode appears in Algorithm 2. We write $x\uparrow$ to denote that the variable $x$ is not defined.

\begin{algorithm} 
\caption{The instructions for process $p_i$ $\textbf{while}\  \mathtt{e}$ is even}  \label{pc:Snowmand}
\begin{algorithmic}[1]
\scriptsize

  \State \textbf{At time $t$:}
  
  \State 
  
  \State \hspace{0.3cm}   \textbf{If} $\mathtt{ready}(\mathtt{e})==0$ \textbf{then} $\mathtt{Init}(\mathtt{e})$, set $\mathtt{ready}(\mathtt{e}):=1$;       \Comment{Initialise values for epoch $\mathtt{e}$}

\State

  \State \hspace{0.3cm} \textbf{If} $\mathtt{newround}==1$:  \label{nrbegind} \Comment Start a new round if ready 

  \State \hspace{0.6cm} Form sample sequence $\langle p_{1,\mathtt{s}},\dots p_{k,\mathtt{s}} \rangle$;  \label{sampled}         \Comment Sample  with replacement

  \State \hspace{0.6cm} For $j\in [1,k]$, send $(\mathtt{s},\mathtt{e})$ to $p_{j,\mathtt{s}}$;  \label{ask}  \Comment Ask  $p_{j,\mathtt{s}}$ for present value

  \State \hspace{0.6cm} Set $\mathtt{start}(\mathtt{s},\mathtt{e}):=t$, $\mathtt{newround}:=0$; \Comment Set $t$ as start time for round $\mathtt{s}$ \label{nrendd}

\State   

\State \hspace{0.3cm} For all $s'\in [0,\mathtt{s}]$ with $t>\mathtt{start}(s',\mathtt{e})\geq t-2\Delta$:  \label{2Deltad} \Comment Update values received 

\State \hspace{0.6cm} For all $j\in [1,k]$: 

\State \hspace{0.9cm} \textbf{If} $\mathtt{rpref}(j,s',\mathtt{e})\uparrow$ and $p_i$ has received a message $(s',B,\sigma,\sigma',\mathtt{e})$ from $p_{j,s'}$ s.t.\ $B$ is a chain
  with $\sigma,\sigma' \subseteq H_B$: 

\State \hspace{1.2cm} Set $\mathtt{rpref}(j,s',\mathtt{e}):=H_B$, $\mathtt{rlock}(j,s',\mathtt{e})=\sigma$, $\mathtt{rfin}(j,s',\mathtt{e}):=\sigma'$; \label{updateendd} \Comment Record values for $p_{j,s'}$ and round $s'$ 

\State \hspace{0.6cm} For  all $\sigma\subseteq \mathtt{pref}$: \label{updatesupportd}

\State \hspace{0.9cm} \textbf{If}  $|\{ j: 1\leq j \leq k, \mathtt{rlock}(j,s',\mathtt{e})\downarrow\supseteq \sigma \}|\geq \alpha_2$:  \label{sfin}

\State \hspace{1.2cm} Set $\mathtt{suppfin}(\sigma,s',\mathtt{e}):=1$; \label{updatesupportendd} \Comment Record that values for round $s'$ support finalizing $\sigma$

\State

\State \hspace{0.3cm} For all $\sigma \subseteq \mathtt{pref}$, \textbf{if} $\mathtt{lock}(\sigma)==0$:  \label{lockupdatestartd}  \Comment Update locks 

\State \hspace{0.6cm} \textbf{If} there exists a least $s'\geq \mathtt{lockbound}(\sigma)$ 
such that

\State \hspace{0.6cm}  $|\{ j: 1\leq j \leq k, \mathtt{rpref}(j,s',\mathtt{e})\supseteq \sigma  \}|\geq \alpha_2 $ \textbf{and} such that

\State \hspace{0.6cm} for all $s'' \in [s',\mathtt{s})$, $\mathtt{pref}(s'',\mathtt{e})\supseteq \sigma$:

\State \hspace{0.9cm} Set $\mathtt{lock}(\sigma):=1$, $\mathtt{lockbound}(\sigma):=s'+1$, $\mathtt{locktime}(\sigma):=t$;




\label{lockupdateendd}

  \State 

   \State \hspace{0.3cm} Set $\mathtt{pref}:=\mathtt{final}$, $\mathtt{end}:=0$; \label{21} \Comment{Begin iteration to update $\mathtt{pref}$} 

  \State \hspace{0.3cm} \textbf{While} $\mathtt{end}==0$ \textbf{do}:  \label{beginwhile}

  \State \hspace{0.6cm} Set $E:= \{ b\in \mathtt{blocks}:\ b \text{ is a child of }\mathtt{last}(\mathtt{pref}) \text{ and }\mathtt{pref} \subseteq \mathtt{reduct}(\mathtt{pref})\ast H(b)\}$; \label{Edef} 

  \State \hspace{0.6cm} \textbf{If} $E$ is empty, set $\mathtt{end}:=1$;

 \State \hspace{0.6cm} \textbf{Else}:        \Comment{Carry out the next instance of Snowflake$^\diamond$}

 \State \hspace{0.9cm} \textbf{If} $\mathtt{val}(\mathtt{pref})$ is undefined:

\State \hspace{1.2cm} Let $b$ be the first block in $E$ enumerated into $\mathtt{blocks}$;  

\State \hspace{1.2cm} Set $\mathtt{val}(\mathtt{pref})$ to be the $(|\mathtt{pref}|+1)^{\text{th}}$ bit of $\mathtt{reduct}(\mathtt{pref})\ast H(b)$;

\State \hspace{0.9cm} \textbf{If} $\mathtt{lock}(\mathtt{pref}\ast \mathtt{val}(\mathtt{pref}))==0$: \label{readybegin}  \Comment Look to see whether we can decide next bit of $\mathtt{pref}(\mathtt{s})$

\State \hspace{1.2cm} \textbf{If} $|\{ j: 1\leq j \leq k, \mathtt{rpref}(j,\mathtt{s},\mathtt{e})\downarrow \not \supseteq \mathtt{pref} \ast (1-\mathtt{val}(\mathtt{pref})) \}| \geq k-\alpha_1+1$: 

\State \hspace{1.5cm} Set $\mathtt{dec}(\mathtt{s},\mathtt{pref} \ast \mathtt{val}(\mathtt{pref} ),\mathtt{e}):=1$;  \label{nr1} 

\State \hspace{1.2cm} \textbf{If} $|\{ j: 1\leq j \leq k, \mathtt{rpref}(j,\mathtt{s},\mathtt{e})\supseteq \mathtt{pref} 
\ast 1-\mathtt{val}(\mathtt{pref}) \}| \geq \alpha_1$: 

\State \hspace{1.5cm} Set $\mathtt{val}(\mathtt{pref}):=1-\mathtt{val}(\mathtt{pref})$;
\State \hspace{1.5cm}  Set $\mathtt{dec}(\mathtt{s},\mathtt{pref} \ast \mathtt{val}(\mathtt{pref} ),\mathtt{e}):=1$;

\State \hspace{0.9cm} \textbf{If} $\mathtt{lock}(\mathtt{pref}\ast \mathtt{val}(\mathtt{pref}))==1$: 

\State \hspace{1.2cm} \textbf{If} $|\{ j: 1\leq j \leq k, \mathtt{rlock}(j,\mathtt{s},\mathtt{e})\downarrow \not \supseteq \mathtt{pref} \ast (1-\mathtt{val}(\mathtt{pref})) \}| \geq k-\alpha_2+1$: 

\State \hspace{1.5cm} Set $\mathtt{dec}(\mathtt{s},\mathtt{pref} \ast \mathtt{val}(\mathtt{pref} ),\mathtt{e}):=1$;

\State \hspace{1.2cm} \textbf{If} $|\{ j: 1\leq j \leq k, \mathtt{rlock}(j,\mathtt{s},\mathtt{e})\supseteq \mathtt{pref} 
\ast 1-\mathtt{val}(\mathtt{pref}) \}| \geq \alpha_2$:  \label{sunlock}

\State \hspace{1.5cm} Set $\mathtt{val}(\mathtt{pref}):=1-\mathtt{val}(\mathtt{pref})$;
\State \hspace{1.5cm}  Set $\mathtt{dec}(\mathtt{s},\mathtt{pref} \ast \mathtt{val}(\mathtt{pref} ),\mathtt{e}):=1$;

\State \hspace{1.5cm} For all $\sigma \supset \mathtt{pref}$, set $\mathtt{lock}(\sigma):=0$ and make $\mathtt{locktime}(\sigma)$ undefined; \label{41} \Comment Unlock

 \State \hspace{0.6cm} Set $\mathtt{pref}:=\mathtt{pref} \ast \mathtt{val}(\mathtt{pref})$; \label{endpref}

\State

\State \hspace{0.3cm} \textbf{If} $\mathtt{start}(\mathtt{s},\mathtt{e})\downarrow \leq t-2\Delta$ \textbf{or} $\mathtt{dec}(\mathtt{s},\sigma,\mathtt{e})==1$ for all  $\sigma$ with $\mathtt{final}\subset \sigma \subseteq \mathtt{pref}$: \label{44}

\State \hspace{0.6cm} Set $\mathtt{pref}(\mathtt{s},\mathtt{e}):=\mathtt{pref}$, $\mathtt{s}:=\mathtt{s}+1$, $\mathtt{newround}:=1$;  \label{45} \Comment Proceed to next round

\label{nr2}

\State 

\State \hspace{0.3cm} \textbf{If} there exists a longest $\sigma$ s.t.\ $\mathtt{final} \subset \sigma \subseteq \mathtt{pref}$ \textbf{and} such that there exists $s'$ such that \textbf{either}: \label{final?}

\State \hspace{0.3cm} (i)  for all $s''\in [s',s'+\beta)$, $\mathtt{suppfin}(\sigma,s'',\mathtt{e})==1$, \textbf{or}: 
\State \hspace{0.3cm} (ii) for all $s''\in [s',s'+1]$,  $|\{ j: 1\leq j \leq k, \mathtt{rfin}(j,s',\mathtt{e})\downarrow  \supseteq \sigma  \}| \geq \alpha_3$: \label{newfin}  \Comment Extra decision condition
\State \hspace{0.6cm} Set  $\mathtt{final}:=\sigma$;  \label{f3} 
\State \hspace{0.6cm} Set  $\mathtt{lastfinalized}:=s'$;       \Comment Records the last round in which new values are finalized

\State

\State \hspace{0.3cm} \textbf{If} $\mathtt{s}-\mathtt{lastfinalized}\geq \gamma$: 
\State \hspace{0.6cm} Send $(\text{stuck},\mathtt{e},\mathtt{final})$ to all processes; \label{stuckm}  \Comment Send epoch $\mathtt{e}+1$ message

\State

\State \hspace{0.3cm} \textbf{If} $p_i$ has received $(s',\mathtt{e})$ from $p_j$ (for the first time):  \label{resps} \Comment Send requested values

\State \hspace{0.6cm} Let $\sigma$ be the longest initial segment of $\mathtt{pref}$ s.t. $t-\mathtt{locktime}(\sigma)\downarrow \geq 4\Delta$; \label{3D}

\State \hspace{0.6cm} Send $(s',\mathtt{chain}(\mathtt{pref}),\sigma,\mathtt{final},\mathtt{e})$ to $p_j$; \label{resps2}

\State 

\State \hspace{0.3cm} \textbf{If} $p_i$ has received an EC for epoch $\mathtt{e}+1$: send the EC to all, set $\mathtt{e}:=\mathtt{e}+1$;  \label{recEC}  \Comment{Enter next epoch}

\end{algorithmic}
\end{algorithm}

\section{Consistency and liveness analysis}  \label{clanalysis} 

We give an analysis for the case that $k=80$, $\alpha_1=41$, $\alpha_2=72$, $\alpha_3=48$, $\beta=14$, $\gamma\geq 300$, and under the assumption that $n\geq 250$ and $f<n/5$. As in \cite{buchwald2025snowman}, we make the assumption that $n\geq 250$ only so as to be able to give as simple a proof as possible: a more fine-grained analysis for smaller $n$ is the subject of future work. To bound the probability of a consistency or liveness failure, we suppose that at most 10,000 processes execute the protocol for at most 1000 years, executing at most 5 rounds per second. 

\subsection{The proof of consistency} \label{consis}

Consistency within even epochs follows from the proof in \cite{buchwald2025snowman}, except that we must account for the new finality condition (line \ref{newfin} of Algorithm 2). To deal with this, we can argue exactly as in \cite{buchwald2024frosty} (where the same finality condition was used). Suppose $p_i$ is correct. A direct calculation for the binomial distribution shows that the probability that at least 3/5 of $p_i$'s sample sequence in a given round are Byzantine is less than $10^{-14}$, i.e., $\text{Bin}(80,0.2,\geq 48)<10^{-14}$. The probability that this happens in two given consecutive rounds is therefore less than $10^{-28}$. Let  $\mathtt{rfin}_i(j,s,e)$  be the value $\mathtt{rfin}(j,s,e)$ as locally defined for $p_i$ at the end of rounds $s$  of epoch $e$, and consider the following event: 
\begin{enumerate}
    \item[$(\diamond_0)$:] If there exists $i,s,e $ and $\sigma$ such that $|\{ j\in [1,k]: \mathtt{rfin}_i(j,s,e)\supseteq \sigma \}|\geq \alpha_3$ and $|\{ j\in [1,k]: \mathtt{rfin}_i(j,s+1,e)\supseteq \sigma \}|\geq \alpha_3$, then some correct processor has already finalized a value extending $\sigma$ by the end of round $s$. 
\end{enumerate}
Conditioned on $(\diamond_0)$, it is not possible for the new decision rule to cause a first consistency violation. So, let us bound the probability that $(\diamond_0)$ fails to hold in a given execution. To achieve this, we can apply  the bound $10^{-28}$ above with the union bound, recalling that we suppose that at most 10,000 processes execute the protocol for at most 1000 years, executing at most 5 rounds per second: $10^{-28} \times 10^{4} \times 5 \times 60 \times 60\times 24 \times 366 \times 10^{3}< 2\times 10^{-13}$. 

\vspace{0.2cm} 
Next, we argue consistency between each even epoch and the following odd epoch, i.e., that if $p_i$ and $p_j$ are correct and if $p_i$ finalizes $\sigma$ during even epoch $e$, while $p_j$ finalizes $\sigma'$ during epoch $e+1$, then $\sigma \subseteq \sigma'$. The proof of  Section 8 of  \cite{buchwald2025snowman} establishes that $(\diamond_1)$ below holds for the entirety of an execution, except with probability at most $2\times 10^{-5}$: 
\begin{enumerate}
    \item[$(\diamond_1)$:] 
Whenever any correct process $p_i$ finalizes some $\sigma$ while in an even epoch $e$,  the following is true of more than  3/4 of the correct processes $p_j$: if $p_j$ ever sends a starting vote $(\text{start}, e+1, \sigma')$, then $\sigma \subseteq \sigma'$. 
\end{enumerate} 

Now suppose that  $(\diamond_1)$ holds, that  $p_i$ and $p_j$ are correct and that $p_i$ finalizes $\sigma$ during even epoch $e$ while $p_j$ finalizes $\sigma'$ during epoch $e+1$. From the definition of the function $\text{fin}$, it follows that $\sigma'$ extends $\mathtt{Pref}(C)$, where $C$ is a starting certificate for epoch $e+1$. From $(\diamond_1)$ and the fact that $f>n/5$ it follows that $\sigma \subseteq \mathtt{Pref}(C)$, so that $\sigma \subseteq \sigma'$, as claimed. 

\vspace{0.2cm} 
Consistency within odd epochs follows as in \cite{chan2023simplex}.  To see this, note that a correct process $p_i$ stops executing an odd epoch $e$ upon receiving a finalized Simplex-blockchain $b_1,\dots,b_{h'}$ for epoch $e$ of reduced height  $\geq \mu$. At this point $p_i$ sets  $\mathtt{final}:=\text{fin}(b_1,\dots,b_{h'})$. So, to ensure consistency within odd epochs, it suffices to ensure that correct processors finalize the same blocks at each height during an odd epoch, which follows directly from the quorum intersection proof in  \cite{chan2023simplex}.

\subsection{The proof of liveness} 
Throughout this section, we assume that the value $E$ (specified in line \ref{Edef} of the pseudocode for even epochs) is never empty for correct $p_i$ when it begins the `while' loop (line \ref{beginwhile}) at the beginning of timeslot $t$, i.e., there are always new blocks to finalize. 

\vspace{0.2cm}
\noindent \textbf{Defining $\mathtt{final}_t$}. At any timeslot $t$, let $\mathtt{final}_t$ be the shortest amongst all local values $\mathtt{final}$ for correct processes (by Section \ref{consis} this value is uniquely defined, except with small  error probability). 

\vspace{0.2cm} The proof of liveness breaks into two parts: 

\vspace{0.1cm} 
\noindent \emph{Claim 3}. Suppose that all correct processes are in even epoch $e$ at $t\geq$GST. Then, except with small  error probability, either $\mathtt{final}_{t+4\Delta \gamma}$ properly extends $\mathtt{final}_t$ or else all correct processes enter epoch $e+1$ by time $t+4 \Delta \gamma$. 

\vspace{0.1cm} 
\noindent \emph{Claim 4}. If all correct processes enter the odd epoch $e+1$, this epoch finalizes a new value. 

\vspace{0.2cm} 
Since Claim 4 follows directly from liveness from Simplex (established in  \cite{chan2023simplex}), it remains to establish Claim 3. 


\vspace{0.2cm}
\noindent \textbf{Establishing Claim 3}. Given the conditions in the statement of the claim, let $P$ be the set of correct processes that have local $\mathtt{final}$ values properly extending $\mathtt{final}_t$ at time $t_1:=t+2\Delta \gamma$. Note that, if $p_i$ is correct then, for any $s$ and $e$, $\mathtt{start}(s+1,e)-\mathtt{start}(s,e)\leq 2\Delta$ (see lines \ref{44}, \ref{45} and lines \ref{nrbegind}--\ref{nrendd}). If $|P|\leq 3n/5$, it follows that at least $n/5$ correct processes send epoch $e+1$ messages $(\text{stuck},e,\mathtt{final}_t)$ by time $t_1$ (see line \ref{stuckm}). In this case, since $t\geq$GST, all correct processes enter epoch $e+1$ by time $t_1 +\Delta$ (see line \ref{recEC}). So, suppose $|P|> 3n/5$ and that it is not the case all correct processes enter epoch $e+1$ by time $t+4 \Delta \gamma$.  Let $x\in \{ 0, 1 \}$ be such that some correct process finalizes $\mathtt{final}_t \ast x$,  (by Section \ref{consis} there exists a unique such $x$ except with small probability). Consider the instructions as locally defined for correct $p_i$ when executing any round $s$ of epoch $e$ that starts subsequent to $t_1$. A calculation for the binomial shows that the probability $|\{ j\in [1,k]: \mathtt{final}(j,s)\supseteq \mathtt{final}_t \ast x \}|\geq 48$ is at least 0.548. 
The probability that this holds in both of any two such consecutive rounds $s$ and $s+1$ is therefore at least 0.3. Since we suppose $\gamma \geq 300$, the probability that $p_i$ fails to finalize a value extending $\mathtt{final}_t$ by time  $t+4 \Delta \gamma$ is therefore at most $0.71^{150}<10^{-22}$. Since we suppose $n<10^4$, it follows from the union bound that the probability $\mathtt{final}_{t+4\Delta \gamma}$ does not properly extend $\mathtt{final}_t$ is $<10^{-18}$.

\bibliographystyle{plainurl}

\end{document}